\documentclass[aps,twocolumn,amssymb,showpacs]{revtex4}

\usepackage{graphicx}

\newcommand{\be}{\begin{equation}}
\newcommand{\ee}{\end{equation}}
\newcommand{\bea}{\begin{eqnarray}}
\newcommand{\eea}{\end{eqnarray}}

\def\k{\kappa}

\begin{document}

\title{On the quantum stress tensor for extreme 2D Reissner-Nordstr\"om black holes}
\author{Roberto Balbinot, Serena Fagnocchi }
\altaffiliation{Email addresses: balbinot@bo.infn.it, fagnocchi@bo.infn.it}
\affiliation{Dipartimento di Fisica dell'Universit\`a di Bologna
and INFN sezione di Bologna, \\  Via Irnerio 46,
40126 Bologna, Italy }
\author{Alessandro Fabbri, Sara Farese and Jos\'e Navarro-Salas}
\altaffiliation{Email addresses: fabbria@bo.infn.it, farese@ific.uv.es, 
jnavarro@ific.uv.es} \affiliation{Departamento
de Fisica Teorica, Facultad de Fisica, Universidad de Valencia, \\
Burjassot-46100, Valencia, Spain}

\begin{abstract}
Contrary to previous claims, it is shown that the expectation values of the
quantum stress tensor for a massless scalar field propagating on a two-dimensional
extreme Reissner-Nordstr\"om black hole  are indeed regular on the horizon.
\end{abstract}


\maketitle

Extremal  black holes play an important role in gravity and string
theories. They appear as soliton like objects with intrinsic
parameters saturating a Bogomol'ni bound and have zero Hawking
temperature \cite{book}. \\ \noindent It is quite disturbing that
quantum field theory (QFT hereafter) on these backgrounds seems to
predict divergences \cite{Trivedi,Hiscock} which, although being
``mild'' in some sense, have no clear physical explanation. These
divergences are associated to ``vacuum'' expectation values of the
stress tensor operator evaluated on the horizon. Before entering
this problem, a digression on the notion of vacuum in this context
is necessary \cite{B&D}.
\\ \noindent In QFT in curved space-time quantization is achieved
as usual by expanding the field operator in annihilation ($a_i$)
and creation ($a_i^{\dagger}$) operators according to a given set
of mode solutions of the field equation. The vacuum, $|0\rangle$,
is the state annihilated by the $a_i$, i.e. $a_i|0\rangle =0$.
However one of the most interesting outcome of this procedure, is
that in the presence of a gravitational field (i.e. in a curved
space-time) the notion of vacuum state becomes rather vague.
Unlike Minkowski QFT, there is no a unique vacuum state. There are
many (in principle infinite) ``vacuum states'', no one sharing the
central and unique role the Minkowski vacuum has for inertial
observers. The ``vacuum states'' one can construct in QFT are not
at all empty (at least everywhere). Furthemore, their particle
content is observer dependent. According to our present
understanding these different vacua simply represent different
physical situations. \\ \noindent In black hole spacetimes one
usually considers three ``vacuum states''.\\  \noindent The first
is the so called Boulware vacuum state $|B\rangle$
\cite{Boulware}. It is constructed with ingoing and outgoing modes
that are positive frequency with respect to the asymptotically
Minkowskian time coordinate. At infinity these modes reduce to
usual plane waves and therefore there the Boulware vacuum reduces
to the Minkowski vacuum ($|B\rangle \sim |0\rangle_M$). If the
behaviour of $|B\rangle$ at infinity seems quite reasonable, the
same cannot be said at the black hole horizon. If the quantum
field is in the Boulware state an inertial observer falling across
the horizon measures an infinite energy density and pressure
\cite{ChristensenFullin}. From the physical point of view
$|B\rangle$ is supposed to describe the vacuum polarization
outside a static star. Being its radius bigger than the horizon,
the above divergence is spurious.
\\ \noindent A quantum state regular on the horizon, the so called
Hartle-Hawking state $|H\rangle$ \cite{IsraelHartleHawking}, can
be constructed using incoming and outgoing modes that are positive
frequency with respect to the affine parameters along the future
and past horizons of the black hole respectively. These ``Kruskal
modes'' do not match at infinity the standard Minkowski plane
waves. At infinity $|H\rangle$ is not empty ($|H\rangle \neq
|0\rangle_M$). It describes equilibrium thermal radiation at the
Hawking temperature $T=\k/2\pi$ (in units $\hbar=c=k_B=1$), where
$\k$ is the surface gravity of the horizon. $|H\rangle$ has the
properties of being a thermal state and is the only static state
which is regular both on the future and past horizons. It is
supposed to describe the thermal equilibrium of a black hole with
its own quantum radiation. Equilibrium is achieved by enclosing
the black hole in a reflecting box. \\ \noindent Finally, the
Unruh state $|U\rangle$ \cite{Unruh} is constructed by ingoing
modes that are positive frequency with respect to the asymptotic
Minkowskian time whereas the outgoing modes are positive frequency
with respect to the affine parameter on the past horizon. This
hybrid construction is from the physical point of view the most
interesting, since it describes the late-time behaviour of a
quantum field in the spacetime of a collapsing body forming a
black hole. From its definition one can show that in this state
one has no particles coming in from past infinity, whereas there
is a thermal flux of particles at the Hawking temperature flowing
out to future infinity. $|U\rangle$ is regular on the future
horizon, but on the past horizon it has the same bad behaviour
$|B\rangle$ had. However for a black hole formed by gravitational
collapse of matter the past horizon does not exist, being covered
by the infalling matter and therefore the related divergence is
spurious.
\\ \noindent
All the features of the ``vacuum states'' we have presented can be
easily seen in a two-dimensional spacetime context where the
quantum field is a massless minimally coupled scalar. For this
case exact analytical results can be found. The spacetime we shall
consider is the 2D section of the Reissner-Nordstr\"om one \be
\label{metric} ds^2=-f(r)dt^2 + \frac{dr^2}{f(r)} =-f(r)dudv \
,\ee where $f(r)=1-2M/r+Q^2/r^2$, $M$ is the mass and $Q$ the
charge of the black hole ($M>|Q|$). $u$ and $v$ are respectively the
retarded and advanced Eddington-Finkelstein coordinates \be
u=t-r^*\ , \ \ \ v=t+r^* \ee where $r^*=\int dr/f(r) $. The
horizon is located at $r_+=M+\sqrt{M^2-Q^2}$. The field equation
for the scalar field is \bea \label{eqpsi}
\Box\psi=0&\Leftrightarrow&\partial_{\bar u}\partial_{\bar v}\psi
=0 \ ,\eea where $\{ \bar u, \bar v \}$ are null coordinates
related to $\{ u,v \}$ by a generic conformal coordinate
transformation $u\to \bar u\ , \ \ v\to \bar v$. The normal modes
of eq.(\ref{eqpsi}) are simply plane waves $\{ e^{-iw\bar u},
e^{-iw\bar v} \} $. Expanding $\psi$ in these modes one constructs
the $|\bar u,\bar v\rangle$ vacuum state. The expectation values of the
quantum stress tensor operator for the $\psi$ field in this state
are \cite{B&D} \bea \label{tmunu} \langle \bar u, \bar
v|T_{uu}|\bar u,\bar v\rangle
&=&-(12\pi)^{-1}f^{\frac{1}{2}}(f^{-\frac{1}{2}})_{,uu}
+\Delta(u,\bar u), \nonumber \\
 \langle \bar u, \bar v|T_{vv}|\bar u,\bar v\rangle &=&-(12\pi)^{-1}f^{\frac{1}{2}}(f^{-\frac{1}{2}})_{,vv}
+\Delta(v,\bar v) ,  \\
\langle T^a_a \rangle &=& (24\pi)^{-1}R= (6\pi)^{-1}f^{-1}(\ln
f)_{,uv} \nonumber \eea where \bea
&&-(12\pi)^{-1}f^{\frac{1}{2}}(f^{-\frac{1}{2}})_{,uu}=-(12\pi)^{-1}f^{\frac{1}{2}}(f^{-\frac{1}{2}})_{,vv}=
 \\
 &&\equiv H(r) = \nonumber \\
&&=(24\pi)^{-1}\bigg(-\frac{M}{r^3}+\frac{3}{2}\frac{M^2+Q^2}{r^4}
-
\frac{3MQ^2}{r^5} + \frac{Q^4}{r^6}\bigg),\nonumber\\
&&(6\pi)^{-1}f^{-1}(\ln
f)_{,uv}=(6\pi)^{-1}\bigg(\frac{M}{r^3}-\frac{3}{2}\frac{Q^2}{r^4}\bigg)
 \eea and \be
\Delta(u,\bar u)=(24\pi)^{-1}\left( \frac{F''}{F}-
\frac{1}{2}\frac{F'^2}{F^2}\right) \ee is the Schwarzian
derivative associated to the transformation $u\to \bar u$ with
$F=du/d\bar u$ and a prime means differentiation with respect to
$u$. Similarly for $\Delta(v,\bar v)$ with $F$ replaced by
$G=dv/d\bar v$. \\ \noindent The last equation in (\ref{tmunu}) is
the well known trace anomaly, where  $R$ is the Ricci scalar. This
expression holds in every conformal vacuum state, and this
explains the omission of the specification of the quantum state.
\\ \noindent Now for the Boulware state $|B\rangle$ the modes are
given by $u=\bar u$, $v=\bar v$, i.e. $\Delta_B(u,\bar
u)=0=\Delta_B(v,\bar v)$ yielding \be \label{boulware} \langle
B|T_{uu}|B\rangle = \langle B|T_{vv}|B\rangle = H(r) \ . \ee For
the Hartle-Hawking state $|H\rangle$, $\bar u=U, \ \bar v=V$,
where $\{ U,V \}$ are the Kruskal coordinates \be \label{Kruskal}
U=-\frac{1}{\k}e^{-\k u}\ , \ \ \ V=\frac{1}{\k}e^{\k v}\ .\ee
$\k$ is the surface gravity at the horizon \be \label{surfgra}
\k=\frac{\sqrt{M^2-Q^2}}{r_+^2}\ . \ee This gives \be
\Delta_H(u,U)=\Delta_H(v,V)=(48\pi)^{-1}\k^2 \ee and \be
\label{hh}\langle H|T_{uu}|H\rangle = \langle H|T_{vv}|H\rangle =
\langle B|T_{uu}|B\rangle + (48\pi)^{-1}\k^2 \ .\ee Finally for
the Unruh state $|U\rangle$, $\bar u=U$, $\bar v=v$ and \bea
\label{unruh} \langle U|T_{uu}|U\rangle &=& \langle
H|T_{uu}|H\rangle =
\langle B|T_{uu}|B\rangle + (48\pi)^{-1}\k^2 \ , \nonumber \\
\langle U|T_{vv}|U\rangle &=& \langle B|T_{vv}|B\rangle \ .\eea
This form of the stress tensor is physically quite interesting
since it is obtained as the late-time behaviour in the case of an
arbitrary collapse in two dimensions. As shown for example in
Birrell and Davies \cite{B&D} the effect of the collapse is to
increase the vacuum polarization part (i.e. $\langle
B|T_{ab}|B\rangle $) with an outgoing (retarded) flux or radiation
that is constant along $u$ rays and that asymptotically approaches
$(48\pi)^{-1}\k^2$.\\ From the previous expressions it is easy to
see that the expectation values in the three states differ just by
conserved traceless radiation at the Hawking temperature
$T_H=\k/2\pi$. Now regularity of the stress tensor (in a regular
frame) on the future horizon is achieved as $r\to r_+$
if \cite{ChristensenFullin} \bea \label{regularity}
f^{-2} T_{uu} &<& \infty\ , \nonumber \\
T_{vv} &<& \infty  \ ,  \\
f^{-1}T_{uv} &<& \infty \nonumber .\eea Regularity on the past
horizon is given by analogous requirements with just $u$ and $v$
interchanged. Using these relations it is easy to verify the
statements made previously concerning the behaviour of the three
quantum states on the horizon.
\\ \noindent
Let us now consider in detail what happens for the extreme
Reissner-Nordstr\"om black hole, for which $M=|Q|$,   since things
become now trickier. This kind of black hole has zero surface
gravity (see eq.(\ref{surfgra})), hence zero Hawking temperature.
This is often stated to imply that the Boulware, Unruh and
Hartle-Hawking states all coincide; for $\k=0$ there is no
difference between the eqs.(\ref{boulware}), (\ref{hh}) and
(\ref{unruh}), namely \bea \langle \ |T_{uu}|\ \rangle &=& \langle
\ |T_{vv}|\ \rangle = -(24\pi)^{-1}
\frac{M}{r^3}\bigg(1-\frac{M}{r}\bigg)^3\nonumber\\
&\equiv& H^{extr}(r)\ ,\label{hext} \\
\langle T^a_a \rangle &=&(24\pi)^{-1}R= (6\pi)^{-1}f^{-1}(\ln
f)_{,uv}\nonumber\\
&=&(6\pi)^{-1}\frac{M}{r^3}\bigg(1-\frac{3}{2}\frac{M}{r}\bigg)
\eea for all three states. Given this, let us check the regularity
conditions on the horizon. It is rather disappointing to see that
the first condition in eqs.(\ref{regularity}) is not satisfied
since \bea \lim_{r\to r_+} f^{-2}\langle T_{uu}\rangle &=& \lim
_{r\to M} \bigg(1-\frac{M}{r}\bigg)^{-4} H^{extr}(r) \nonumber
\\
&&=-(24\pi)^{-1}\frac{M}{r^3}\bigg(1-\frac{M}{r}\bigg)^{-1} \eea
diverges. A similar divergence is found in the past horizon. An
observer crossing the horizon measures therefore an unbounded
energy density and pressure. It must be noted, however, 
that this singularity is regarded to be
suficiently ``mild'' since it leads to finite tidal distortions
and finite curvature \cite{Trivedi}. In any case, before reaching any
definitive conclusion one should take the extreme black hole limit
with care since the Kruskal coordinates transformation given by
eq.(\ref{Kruskal}) makes no sense for $\k=0$ and hence the
expression eq.(\ref{hext}), obtained by calculating the Schwarzian
derivative and taking the limit $\k=0$ at the end, becomes rather
doubtful. \\
On the other hand, despite the mathematical inconsistency of its
derivation, the stress tensor whose components are given by
eq.(\ref{hext}) is the only conserved tensor in the extreme 2d
Reissner-Nordstr\"om spacetime with the correct trace anomaly
which is static (i.e. has the time translation invariance of the
underlying manifold) and vanishes asymptotically as a zero
temperature equilibrium state should do. The solidity of this
argument seems to leave no way out concerning the singular
behaviour at the horizon.\\
The critical point is whether the state whose stress tensor is
given by eq.(\ref{hext}) has any physical significance, i.e. it
can be realized by some physical process.\\ To examine this point,
let us consider the formation of an extremal black hole and the
correspondingly stress tensor for a massless scalar field
propagating
in this geometry.\\
For simplicity let us model the collapsing body forming the black
hole by an ingoing null shell. The space time is depicted in Fig.1
\begin{figure}
\includegraphics[angle=0,width=3.5in,clip]{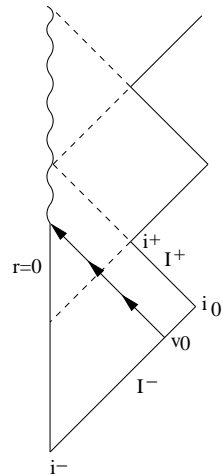}
\caption{Penrose diagram describing the formation of an extreme
Reissner-Nordstr\"om black hole.} \label{fig1}
\end{figure}
An incoming null shell at $v=v_0$ creates an extreme
Reissner-Nordstr\"om black hole whose metric is given by: \bea
ds^2&=&-\bigg(1-\frac{M}{r}\bigg)^2dt^2+\bigg(1-\frac{M}{r}\bigg)^{-2}dr^2\\
&=&-\bigg(1-\frac{M}{r}\bigg)^2dudv \nonumber \eea where as usual
$u=t-r^\ast$, $v=t+r^\ast$ but now: \be r^\ast=\int
\frac{dr}{\big(1-\frac{M}{r}\big)^2}=r+2M\ln\bigg(\frac{r}{M}-1\bigg)-\frac{M^2}{(r-M)}.
\ee In the past of the shell the spacetime is Minkowski, with
metric \be ds^2=-dT^2+dr^2=-d\bar u d\bar v \ee and \bea \bar u
&=& T-r ,\nonumber\\\bar v &=& T+r.\eea Asymptotic flatness in the
past ($\mathcal{I}^{-}$ exists) implies $v=\bar v$.\\
An incoming $\bar v$ mode from past infinity is reflected from the
regular Minkowski origin $r=0$ (i.e. $\bar u=\bar v$) and becomes
an outgoing $\bar u$ mode. Matching across the shell at $v=\bar
v=v_0$ yields \be u=\bar u-4M \bigg[\ln\big(\frac{v_0-\bar
u}{2M}-1\big)-\frac{1}{2\big(\frac{v_0-\bar u}{2M}-1\big)}\bigg].
\ee Being the horizon ($r=M$) located at $\bar u=v_0-2M$, i.e.
$u\to+\infty$, it is easy to see that a positive frequency $\bar
v$ mode on $\mathcal{I}^{-}$ becomes at late advanced time
($u\to+\infty$) a positive frequency $U$ mode, where \be u=-4M
\bigg[\ln\bigg(-\frac{U}{M}\bigg)+\frac{M}{2U}\bigg] \label{u} \ee
and we have redefined $\bar u\to 2U-2M+v_0$ so that the horizon
lies now at $U=0$.\\
The relation (\ref{u}) has been proposed by Liberati et al.
\cite{Liberati} as generalization of the Kruskal transformation to
the extreme Reissner-Nordstr\"om black hole. Gao \cite{Gao} has
later shown that eq.(\ref{u}) defines indeed a smooth extension
across the horizon. To get this result, the presence of the
subleading (as $U\to 0$) logarithmic term is critical.\\
One can
now evaluate in the future of the shell the stress tensor using
the general expression of eq.(\ref{tmunu}). The late time
behaviour defines the Unruh state ($\bar u=U,\bar v=v$) for the
extreme Reissner-Nordstr\"om black hole \bea \langle U|T_{uu}|U
\rangle &=& H^{extr}(r)+ \Delta(u,U), \nonumber\\
\langle U|T_{vv}|U \rangle &=&H^{extr}(r),\\
\langle T^a_a \rangle &=& (24\pi)^{-1}R= (6\pi)^{-1}f^{-1}(\ln
f)_{,uv}. \nonumber \eea The Schwarzian derivative $\Delta(u,U)$
calculated from eq.(\ref{u}) yields \be \Delta(u,U)=
(24\pi)^{-1}\frac{U^3(U-2M)}{2M^2(2U-M)^4},\ee so that \bea
\label{tUU} \langle U|T_{uu}|U \rangle
&=&-(24\pi)^{-1}\frac{M}{r^3}\bigg(1-\frac{M}{r}\bigg)^3+\nonumber\\
&&+(24\pi)^{-1}\frac{U^3(U-2M)}{2M^2(2U-M)^4}, \\
\langle U|T_{vv}|U \rangle &=&-(24\pi)^{-1}\frac{M}{r^3}\bigg(1-\frac{M}{r}\bigg)^3,\nonumber\\
\langle T^a_a \rangle
&=&(6\pi)^{-1}\frac{M}{r^3}\bigg(1-\frac{3}{2}\frac{M}{r}\bigg),
\nonumber \eea which when compared with the expression obtained by
naively taking the limit $\k=0$ (eq.(\ref{hext})) shows a striking
difference in the $T_{uu}$ component. Note that the tensor of
eq.(\ref{tUU}) is conserved, has the correct trace but is not time
independent. So
there is no contraddiction with our previous remark on the unicity of the expression eq.(\ref{hext}).\\
The most remarkable feature of the stress tensor (\ref{tUU}) is
that it is regular on the future event horizon. Using $U=-(r-M)$
one can easly check that \bea && \lim_{r\to M} f^{-2}\langle
U|T_{uu}|U \rangle=\nonumber\\
&=&\lim_{r\to M}
-\frac{1}{24\pi}\bigg[\frac{M}{r^3}\bigg(1-\frac{M}{r}\bigg)^{-1}-\frac{1}{f^2}\Delta(u,U)\bigg]\nonumber\\
&=&-\lim_{r\to
M}\frac{1}{24\pi}\bigg[\frac{M}{r^2}\frac{1}{r-M}-\frac{1}{M(r-M)}+finite
\bigg]\nonumber\\
&=&-\frac{1}{24\pi}\bigg(\frac{3}{2M^2}\bigg)<\infty.\eea The
divergence in $f^{-2}\langle U|T_{uu}|U \rangle$ coming from the
vacuum polarization part (i.e. $H^{extr}(r)$) is exactly canceled
by the divergent term $f^{-2}\Delta(u,U)$.\\
One should appreciate the fundamental role played by the
subleading logarithmic term in the relation between $u$ and $U$
(eq.(\ref{u})). Omission of this term (which corresponds to the
extension proposed by Lake \cite{Lake}) yields an identically
vanishing Schwarzian derivative and the resulting stress tensor
would reduce again to the static one of eq.(\ref{hext}) with the
associated divergence on the horizon. However, as already
stressed, the logarithmic term is necessary to have a smooth
extension across horizon and this explains the regular behaviour
of the stress tensor (eq.(\ref{tUU})) on the horizon which emerges
from our analysis. This result is not a peculiar feature of the
simple collapse model (null shell) we have used. The asymptotic
relation eq.(\ref{u}) is completely general in two dimensions
\cite{saraserena} and its validity can be extended to the physical
spacetime (i.e. four dimensions) because of the propagation of
outgoing rays near the horizon according to geometric optics \cite{Gao}. The
late time radiation is indipendent of the details of the collapse
which affects the $O(U^4)$ term but not the $O(U^3)$. The stress
tensor of eq.(\ref{tUU}) is explicitly time dependent.
Asymptotically (i.e. $r\to \infty$) there is no incoming radiation
on $\mathcal{I}^{-}$ whereas on $\mathcal{I}^{+}$ there is an
outgoing flux given by $\Delta (u,U)$ vanishing at late advanced
time ($U\to 0$,$u\to+\infty$). However, unlike the $\k\neq 0$ case
one can not simply discard this radiation term to get the late
time behaviour since this procedure would lead to the incorrect
result of eq.(\ref{hext}). Also the vacuum polarization part is
vanishing for $U\to 0$ and a careful consideration of both
terms is required as we have shown to have regularity on the future horizon.\\
Finally, one can consider both the future and past extensions
across the horizon, namely, introducing Kruskal like coordinates
$(U,V)$ \cite{Liberati}, \bea u&=&-4M
\bigg[\ln\big(-\frac{U}{M}\big)+\frac{M}{2U}\bigg],\\
 v&=&4M
\bigg[\ln\big(\frac{V}{M}\big)-\frac{M}{2V}\bigg], \nonumber\eea
and define a state which is regular both on the future and past
horizons \bea \langle H|T_{uu}|H \rangle &=&
H^{extr}(r)+\Delta(u,U)= \langle U|T_{uu}|U \rangle, \nonumber\\
\langle H|T_{vv}|H \rangle &=& H^{extr}(r)+\Delta(v,V)=\\
&=&\langle U|T_{vv}|U \rangle +(24\pi)^{-1}\frac{V^3(V+2M)}{2M^2(2V+M)}, \nonumber\\
\langle \ |T_{a}^{a}|\ \rangle&=&(24\pi)^{-1}R=
(6\pi)^{-1}f^{-1}(\ln
f)_{,uv}=\nonumber \\
&=&(6\pi)^{-1}\frac{M}{r^3}\bigg(1-\frac{3}{2}\frac{M}{r}\bigg).\eea
This state is by no way unique. Any smooth extension of the
coordinates $(U,V)$ will lead to the same $U^3$ ($V^3$) behaviour in the
Schwarzian derivative on the horizon responsible for the
divergence cancellation, the difference being of order $U^4$ ($V^4)$.


\end{document}